\documentclass[12pt,a4paper]{article}

\usepackage[top=3cm, bottom=3cm, left=2.5cm, right=2.5cm]{geometry}

\usepackage{amsmath,amssymb}
\usepackage{comment}
\usepackage{txfonts}
\usepackage{cite}
\usepackage{arydshln}
\usepackage{caption}
\usepackage[dvipdfmx]{graphicx}
\usepackage{hyperref}
\usepackage[T1]{fontenc}
\usepackage{here}
\hypersetup{
	colorlinks=true,
	linkcolor=red,
	citecolor=blue
}
\newcommand{\one}{\boldsymbol{1}}
\newcommand{\zero}{\boldsymbol{0}}
\newcommand{\Tr}{\mathrm{Tr}}

\makeatletter

\@addtoreset{equation}{section}
\makeatother

\begin{document}
\title{
	{\Large \bf  Instantons and Berry's connections on quantum graph}
	\\*[10pt]
}

\author{
	Tomonori Inoue\footnote{E-mail: {\url{t-inoue@stu.kobe-u.ac.jp}}}\ ,\ \ \ 
	Makoto Sakamoto\footnote{E-mail: \url{dragon@kobe-u.ac.jp}}\ , \ and\ \ 
	Inori Ueba\footnote{E-mail: \url{i-ueba@stu.kobe-u.ac.jp}}\\*[10pt]
	{\it  Department of Physics, Kobe University, Kobe 657-8501, Japan}\\*[30pt]
}

\date{}

\begin{titlepage}
		\maketitle
		\thispagestyle{empty}

	\begin{flushright}
			\vspace{-8cm}KOBE-TH-21-01
			\vspace{7cm}\\
	\end{flushright}

	\abstract{
		\normalsize
			In this paper, we study non-Abelian
			Berry's connections in the parameter space of boundary conditions for Dirac zero modes on quantum graphs.
			We apply the ADHM construction, which is the method for constructing Yang-Mills instanton solutions, to the Berry's connections. 
			Then we
			find that the instanton configurations appear as the
			Berry's connections.
		}	
	\begin{flushleft}
\end{flushleft}
\end{titlepage}

\section{Introduction}
\label{sec:Introduction}
A quantum graph is known as a quantum mechanical system on a one dimensional graph which consists of edges and vertices connected with each other. The graph is characterized by
boundary conditions (or connection conditions) imposed on each vertex for wavefunctions by the requirement of the probability conservation. (For reviews of quantum graph, see~\cite{Kuchment_2004,Kuchment_2005}.)
This system has been applied to various research areas such as scattering theory, nanotechnology on one dimensional graphs~\cite{Kostrykin:1998gz,Texier_2001,Boman:2005,Fujimoto:2018lzq}, quantum chaos~\cite{Kottos:1997,Cheon:2006,Gnutzmann:2010}, anyons~\cite{Harrison_2014,Maciazek:2017jon,Maci_ek_2019}, supersymmetric quantum mechanics~\cite{Nagasawa:2003tw,Nagasawa:2005kv,Ohya:2012qz}, extra dimensional models~\cite{Fujimoto:2012wv,Fujimoto:2014fka,Fujimoto:2019fzb} and so on, due to fascinating structures from boundary conditions.
Therefore, the study of boundary conditions on the quantum graph will contribute to the further development of low and high energy physics. 

One of features of quantum graphs is the existence of degeneracies in the energy spectrum depending on boundary conditions that respect certain symmetries or topological properties.
The degeneracies will lead to non-Abelian Berry's connections~\cite{Wilczek:1984dh} by considering adiabatic changes of parameters of boundary conditions. For example, 
Refs.~\cite{Ohya:2014ska,Ohya:2015xya} studied non-Abelian Berry's connections in the case of simple boundary conditions and showed that configurations of non-Abelian monopoles appear.
Since a general quantum graph has a larger parameter space, we can expect that other non-trivial configurations of Berry's connections exist on the parameter space of boundary conditions.

In this paper, we investigate non-Abelian Berry's connections in a parameter space of  boundary conditions of quantum graphs to reveal further nontrivial configurations. 
In the previous paper~\cite{Fujimoto:2019fzb}, we have considered a 1+4 dimensional (1+4d) Dirac fermion on quantum graphs and classified the degeneracy of Dirac zero modes.
This study will also be applicable to a fermion on quantum graphs in various dimensions.
Therefore, for simplicity, we will focus on the case of a 1+1d Dirac fermion on quantum graphs which would be realized and controllable by a future development of the nanotechnology, and consider Berry's connections for its degenerate zero modes.
Surprisingly, we find that the structure of the Atiyah--Drinfeld--Hitchin--Manin (ADHM) construction~\cite{Atiyah:1978ri} is hidden in the Berry's connections, which is known as the method of constructing general instanton solutions of non-Abelian gauge theories on the 4d Euclidean space.
Then, by applying the ADHM construction, we will show that the instantons with nontrivial topological charges can be obtained as the non-Abelian Berry's connections.

This paper is organized as follows: In the next section, we briefly review the ADHM construction of {\it SU(N)} instantons. In section \ref{sec:boundary-conditions-and-zero-modes-on-quamtum-graph},
we classify allowed boundary conditions and examine Dirac zero modes on quantum graphs.
Then, in section \ref{sec:berrys-connection}, we consider Berry's connections for Dirac zero modes. We show that the ADHM construction can be applied to Berry's connections on boundary conditions
and they
can be given by the configuration of instantons.
We also see some concrete examples.
The section \ref{sec:conclusion-and-discussion} is devoted to
conclusion and discussion.

\section{Instantons and the ADHM construction}
\label{sec:instantons-and-the-adhm-construction}
In this section, we briefly review the ADHM construction of instantons for ${\it SU(n)}$ Yang--Mills theory. For the detail of the construction, see \cite{Corrigan:1983sv}.
The ones for the $O(n)$ and ${\it Sp}(n)$ gauge group are also discussed in \cite{Christ:1978jy}.

The (anti-)self-dual equation of ${\it SU(n)}$ Yang--Mills theory on the 4d Euclidean space is given by
	\begin{align}
		F_{\mu\nu}=\pm\tilde F_{\mu\nu}\ \ \ \ (\mu,\nu=1,2,3,4)\,.
		\label{eq:self_dual_equation}
	\end{align}
Here the field strength $F_{\mu\nu}$ and its dual $\tilde F_{\mu\nu}$ are defined as
	\begin{align}
		F_{\mu\nu}=\partial_\mu A_\nu-\partial_\nu A_\mu+[A_\mu,A_\nu]\,,
		\ \ \ \ 
		\tilde F_{\mu\nu}=\frac{1}{2}\epsilon_{\mu\nu\rho\lambda}F_{\rho\lambda}\,,
	\end{align}
and we take the gauge field $A_\mu$ as an $n\times  n$ anti-Hermitian matrix.
Instantons are the solutions of Eq.~\eqref{eq:self_dual_equation} and classified by the topological charge, {\it i.e.} the second Chern number
	\begin{align}
		Q=-\frac{1}{16\pi^2}\int d^4x\,\Tr(F_{\mu\nu}\tilde F_{\mu\nu})\,.
		\label{eq:instanton_number}
	\end{align}

Let us then consider the ADHM construction. As an example, we discuss the case for anti-self-dual instantons with topological charge $-k\ (k>0)$.
First, we introduce the matrix $\Delta(x)$
	\begin{align}
		\Delta_{[n+2k]\times[2k]}(x)=a_{[n+2k]\times[2k]}+b_{[n+2k]\times[2k]}\cdot(x^\mu e_\mu\otimes \one_k )\,,
	\end{align}
where the subscript $[l]\times[m]$ indicates the size of $l\times m$ matrix.  
$a$ and $b$ are $(n+2k)\times 2k$ complex matrices which are called the ADHM data and $e_\mu=(-i\sigma^i,\one_2)$. 
Furthermore,
we require the constraints that $\Delta^\dagger \Delta$ is invertible for all $x^\mu$ and commutes with the Pauli matrices $\sigma^i\otimes\one_k$\,:
	\begin{align}
		&\Delta^\dagger\Delta\,f=\one_{2k}\,,
		\label{eq:ADHM_invertible}
		\\
		&[\Delta^\dagger\Delta\ ,\ \sigma^i\otimes\one_k]=0\ \ \ \ \ (i=1,2,3)\,.
		\label{eq:ADHM_constraint}
	\end{align}
The existence of the inverse matrix $f$ is equivalent to the linearly independence of the $2k$ column vectors in $\Delta$.

Second, we consider the $(n+2k)\times n$ matrix $v(x)$, whose $n$ column vectors are orthonormal with each other and span the complementary space for the vectors in $\Delta$. The matrix $v(x)$ is required to satisfy the zero mode equation and the normalization condition
	\begin{align}
		&\Delta^\dagger v(x)=0\,,
		\label{eq:Dirac_equation_V}
		\\
		&v^\dagger v=\one_n\,.
		\label{eq:normaization_V}
	\end{align}
Then we can construct the anti-self-dual instanton solutions with the topological charge $Q=-k$ as follows:
	\begin{align}
		A_\mu(x)=v^\dagger(x)\,\partial_\mu v(x)\,.
		\label{eq:instanton}
	\end{align}
We can also obtain self-dual instantons by using $e^\dagger_\mu$ instead of $e_\mu$ in the above discussion.

The transformation $v(x)\to  v(x)g(x)\ \ (g(x)\in {\it SU}(n))$ gives a gauge transformation of $A_\mu$
	\begin{align}
		A_\mu(x)\to g(x)^{-1}A_\mu(x)g(x)+g(x)^{-1}\partial_\mu g(x)\,.
	\end{align}
Furthermore, we find that the following transformation does not change the conditions \eqref{eq:ADHM_constraint}, \eqref{eq:Dirac_equation_V}, \eqref{eq:normaization_V}
and leads to the same instanton solutions:
	\begin{align}
		a\to \mathcal{Q}\,a\,\mathcal{R}\,,\ \ \ \ \ \ 
		b\to \mathcal{Q}\,b\,\mathcal{R}\,,\ \ \ \ \ \ 
		v(x)\to \mathcal{Q}\,v(x)\,\,,
	\end{align}
where $	\mathcal{Q}\in U(n+2k)\,,\ \ \mathcal{R}\in \one_2\otimes GL(k;\mathbb{C})$.
From this transformation, we can put the ADHM data into what is called the canonical form and fix the degrees of freedom of $b$,
	\begin{align}
		a=\begin{pmatrix}
		\ S_{[n]\times[2k]}\  
		\\
		-T_{[2k]\times[2k]}
		\end{pmatrix}
		=\begin{pmatrix}
		\ \ I^\dagger_{[n]\times[k]} \ \ \  J_{[n]\times[k]}\ \ 
		\\
		\ -e_\mu\otimes T^\mu_{[k]\times[k]}\ 
		\end{pmatrix}
		\,,
		\ \ \ \ \ \ 
		b=\begin{pmatrix}
		\ \zero_{[n]\times[2k]}\ 
		\\
		\ \one_{[2k]\times[2k]}\ 
		\end{pmatrix}\,.
		\label{eq:canonical_form}
	\end{align}
$I^\dagger,J$ are $n\times k$ complex matrices and $T^\mu$ are given by $k\times k$ Hermitian matrices to satisfy \eqref{eq:ADHM_constraint}.\footnote{The canonical form has residual symmetries 
	\begin{align}
		a\to\begin{pmatrix}
		Q & \zero\\
		\zero & R^\dagger
 		\end{pmatrix}a
 		R\,,
 		&&
			b\to\begin{pmatrix}
		Q & \zero\\
		\zero & R^\dagger
		\end{pmatrix}b
		R=b\,,
		&& Q\in U(n)\,,\ R\in \one_2\otimes U(k)\,.
		\notag
	\end{align}
}
\ In this form, the constraint \eqref{eq:ADHM_constraint} can be rewritten into the so-called ADHM equation
	\begin{align}
		[T^1,T^2]+[T^3,T^4]-\frac{i}{2}(I I^\dagger-J^\dagger J)=0\,,
		\\
		[T^1,T^3]-[T^2,T^4]-\frac{1}{2}(IJ-J^\dagger I^\dagger)=0\,,
		\\
		[T^1,T^4]+[T^2,T^3]-\frac{i}{2}(IJ+J^\dagger I^\dagger)=0\,.
	\end{align}

In Section \ref{sec:berrys-connection}, we will see that the structure of the ADHM construction naturally appears in the Berry's connections on the parameter space of boundary conditions of quantum graphs.

\section{Boundary conditions and zero modes on quantum graph}
\label{sec:boundary-conditions-and-zero-modes-on-quamtum-graph}

\begin{figure}[h]
	\centering
	\includegraphics[width=80mm]{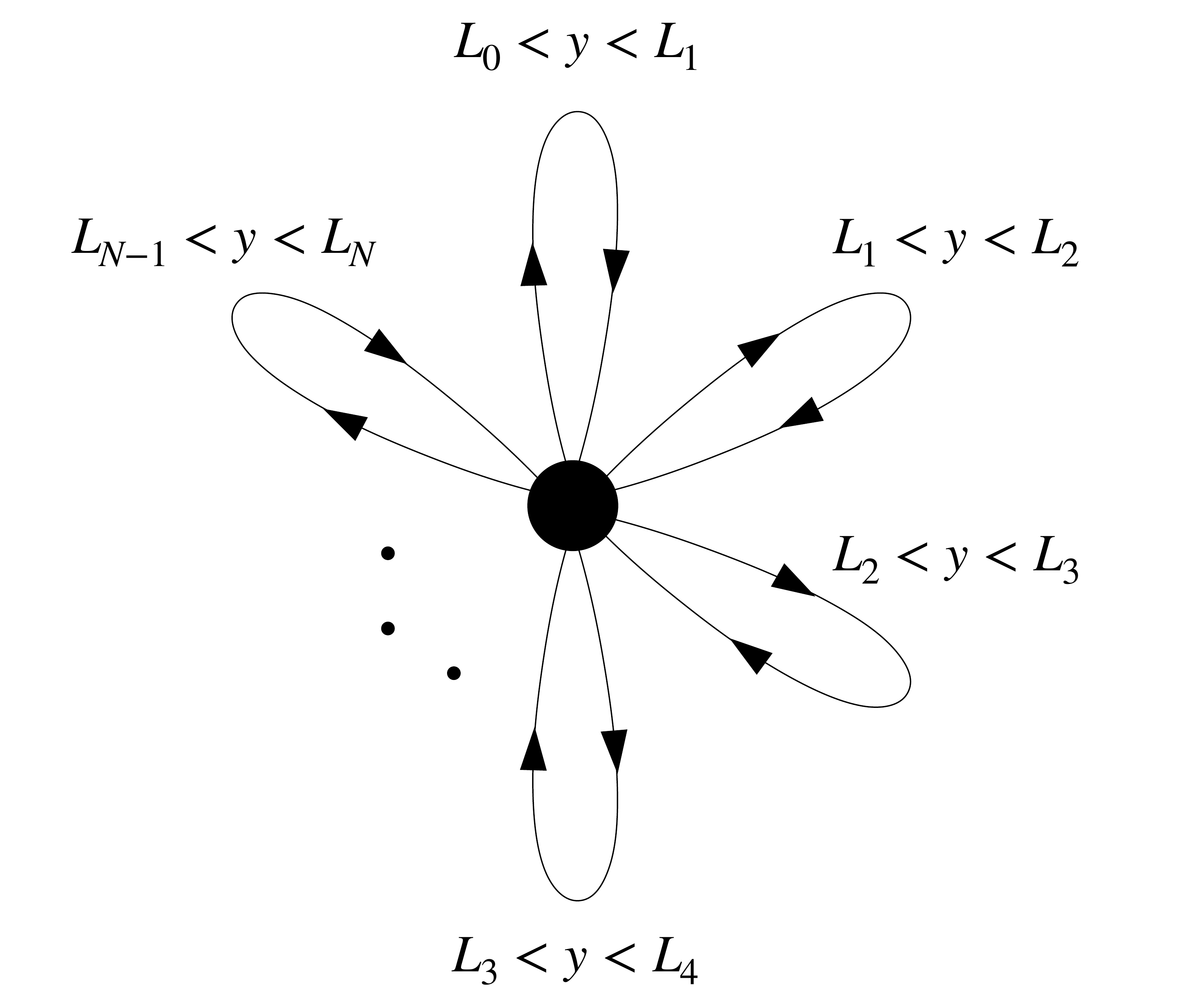}
	\vspace{-0.5cm}
	\caption{A rose graph with one vertex and $N$ loops.}
	\label{fig:rose_graph}
\end{figure}

Let us discuss a 1+1d fermion on a quantum graph. Degenerate zero modes can appear in this system.
As a quantum graph, we consider a rose graph consisting of one vertex and {$N$} edges, each of which forms a loop with $L_{a-1}<y<L_{a}$ ($a=1,2,\cdots,N$): see Fig.~\ref{fig:rose_graph}.

\begin{figure}[t]	
	\begin{center}
		\begin{minipage}{0.4\hsize}
				\centering
				\includegraphics[width=60mm]{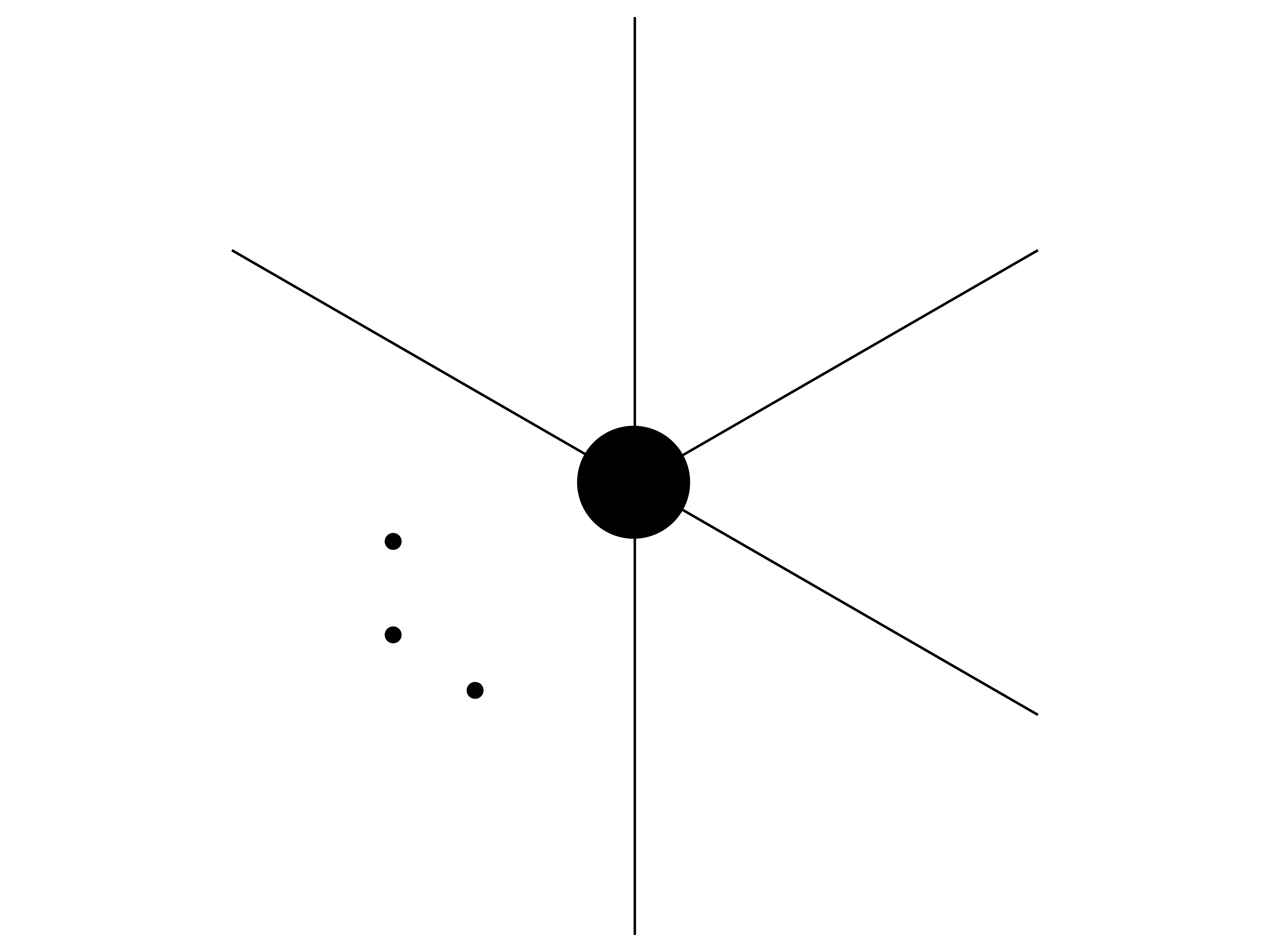}
				\caption{Star graph}
				\label{fig:star-graph}
		\end{minipage}
	\hspace{3cm}
		\begin{minipage}{0.4\hsize}
				\includegraphics[width=60mm]{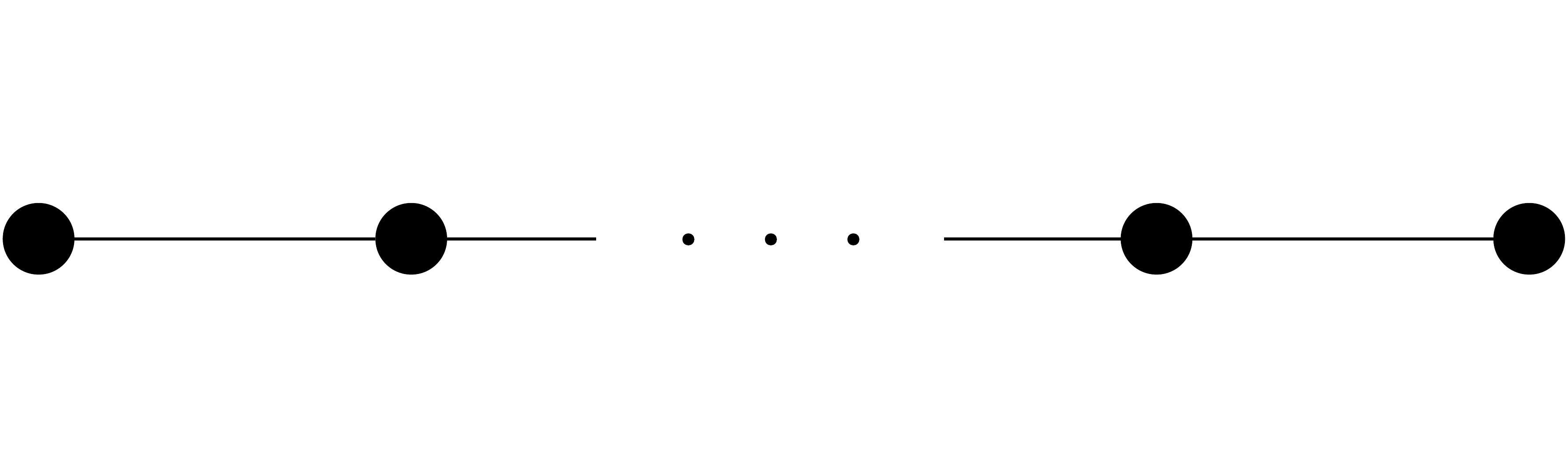}
				\caption{Interval with point interactions}
				\label{fig:point_interaction-graph}
		\end{minipage}
	\end{center}
\end{figure}

It should be noted that the rose graph can be regarded as a master quantum graph because this graph can reduce to arbitrary graphs with the same number of edges, e.g.
a star graph (Fig.~\ref{fig:star-graph}), an interval with point interactions (Fig.~\ref{fig:point_interaction-graph}) and so on, by appropriately tuning boundary conditions at the vertex. 

\subsection{1+1d fermion on quantum graph}

The 1+1d Dirac equation is given by
	\begin{align}
		\left[i\gamma^0\partial_t+i\gamma^1\partial_y-M\right]\Psi(t,y)=0\,,
		\label{eq:Dirac_equation}
	\end{align}
or equivalently
	\begin{align}
		&i\partial_t\,\Psi(t,y)=\mathcal{H}_D\,\Psi(t,y)\,,
		\ \ \ \ &\mathcal{H}_D=-i\gamma^0\gamma^1\partial_y+\gamma^0M\,,
	\end{align}
where $\Psi(t,y)=(\psi_1(t,y),\,\psi_2(t,y))^{\rm T}$ is a two-component spinor, $M$ denotes a mass and $\mathcal{H}_D$ corresponds to the Dirac Hamiltonian.
$\gamma^\mu\ (\mu=0,1)$ indicate $2\times 2$ gamma matrices, whose representation in this paper are defined as 
	\begin{align}
		\gamma^0=\sigma^1\,,\ \ \ \ \gamma^1=i\sigma^3\,.
		\label{eq:gamma_rep}
	\end{align}	
	
From the probability conservation or equivalently the Hermiticity of $\mathcal{H}_D$, we require that the probability current $j=\Psi^\dagger\gamma^0\gamma^1\Psi$ satisfies the following relation at the vertex:
	\begin{align}
		\sum_{a=1}^N \Big[j(t,y)\Big]_{y=L_{a-1}+\varepsilon}^{y=L_{a}-\varepsilon}=	-i\sum_{a=1}^N \Big[\psi_1^\dagger(t,y)\psi_2(t,y)-\psi_2^\dagger(t,y)\psi_1(t,y)\Big]_{y=L_{a-1}+\varepsilon}^{y=L_{a}-\varepsilon}=0\,,
		\label{eq:provability_conservation}
	\end{align}
where $\varepsilon$ is an infinitesimal positive constant.
Since $j$ can be also regarded as the electric current,
the above equation physically corresponds to the Kirchhoff's rule in electromagnetism at the vertex and ensures the charge conservation on the quantum graph. (See also \cite{Kostrykin:1998gz}.)
If we introduce the $2N$-component boundary vectors
	\begin{align}
		\vec F(t)=
		\begin{pmatrix}
		\psi_1(t,L_0+\varepsilon)
		\\
		\psi_1(t,L_1-\varepsilon)
		\\
		\psi_1(t,L_1+\varepsilon)
		\\
		\psi_1(t,L_2-\varepsilon)
		\\
		\vdots
		\\
		\psi_1(t,L_{N-1}+\varepsilon)
		\\
		\psi_1(t,L_N-\varepsilon)
		\end{pmatrix}
		\,,
		\ \ \ \ 
		\vec G(t)=
		\begin{pmatrix}
		\psi_2(t,L_0+\varepsilon)
		\\
		-\psi_2(t,L_1-\varepsilon)
		\\
		\psi_2(t,L_1+\varepsilon)
		\\
		-\psi_2(t,L_2-\varepsilon)
		\\
		\vdots
		\\
		\psi_2(t,L_{N-1}+\varepsilon)
		\\
		-\psi_2(t,L_N-\varepsilon)
		\end{pmatrix}
		\,,
	\end{align}
Eq. \eqref{eq:provability_conservation} can be written into the following form with an arbitrary nonzero real constant $\kappa$
	\begin{align}
		|\vec F(t)+i\kappa \vec G(t)|^2=|\vec F(t)-i\kappa \vec G(t)|^2\,.
	\end{align}
This implies
the norms of the left and right-hand side vectors are equivalent, and therefore we can obtain the boundary condition characterized by a $2N\times 2N$ unitary matrix $U$:
	\begin{align}
		(\one_{2N}-U)\vec F(t)=i\kappa (\one_{2N}+U)\vec G(t)\,.
		\label{eq:general_BC}
	\end{align}

Next, let us consider the energy eigenfunctions $\Psi_{n}^{(i)}(y)=(f_n^{(i)}(y),g_n^{(i)}(y))^\mathrm{T}$ which satisfy
	\begin{align}
		\mathcal{H}_D
		\Psi_{n}^{(i)}(y)
		=E_n
		\Psi_{n}^{(i)}(y)\,,
		\label{eq:Dirac_eq_mode_function}
	\end{align}
or equivalently
	\begin{align}
		&(\partial_y+M)f_n^{(i)}(y)=E_ng_n^{(i)}(y)\,,
		\label{eq:Dirac_eq_f}
		\\
		&(-\partial_y+M)g_n^{(i)}(y)=E_nf_n^{(i)}(y)\,,
		\label{eq:Dirac_eq_g}
	\end{align}
where $E_n$ are the energy eigenvalues. 
The index $n$ denotes the level of energies and $i$ is the label which distinguishes the degeneracy of each energy eigenstate. Positive (negative) energies are labeled by $n>0\ (n<0)$ and $n=0$ corresponds to the zero energy. 
We take the mode functions to be orthonormal
	\begin{align}
		\sum_{a=1}^{N}\int_{L_{a-1}}^{L_a}dy\,(\Psi_{n}^{(i)}(y))^\dagger\Psi_{n'}^{(i')}(y)=\delta_{nn'}\delta^{ii'}\,.
	\end{align} 
	
By means of these eigenfunctions, we can decompose the wavefunction as 
	\begin{align}
		\Psi(t,y)=
		\sum_{n,\,i}a_n^{(i)}e^{-iE_nt}\Psi_{n}^{(i)}(y)\,,
		\label{eq:decomposition_psi}
	\end{align}
where $a_n^{(i)}$ indicate arbitrary coefficients. Then we obtain the boundary condition for the energy eigenfunctions:
	\begin{align}
		(\one_{2N}-U)\vec F_n^{(i)}=i\kappa (\one_{2N}+U)\vec G_n^{(i)}
		\ \ \ \ \ \text{for} \ \ \ \ \ {}^\forall n\ \text{and}\ {}^\forall i\,.
		\label{eq:BC_mode_function}
	\end{align}
Here we also introduced the $2N$-component boundary vectors for the energy eigenfunctions
	\begin{align}
		\vec F_n^{(i)}=
		\begin{pmatrix}
		f_n^{(i)}(L_0+\varepsilon)
		\\
		f_n^{(i)}(L_1-\varepsilon)
		\\
		f_n^{(i)}(L_1+\varepsilon)
		\\
		f_n^{(i)}(L_2-\varepsilon)
		\\
		\vdots
		\\
		f_n^{(i)}(L_{N-1}+\varepsilon)
		\\
		f_n^{(i)}(L_N-\varepsilon)
		\end{pmatrix}
		\,,
		\ \ \ \ 
		\vec G_n^{(i)}=
		\begin{pmatrix}
		g_n^{(i)}(L_0+\varepsilon)
		\\
		-g_n^{(i)}(L_1-\varepsilon)
		\\
		g_n^{(i)}(L_1+\varepsilon)
		\\
		-g_n^{(i)}(L_2-\varepsilon)
		\\
		\vdots
		\\
		g_n^{(i)}(L_{N-1}+\varepsilon)
		\\
		-g_n^{(i)}(L_N-\varepsilon)
		\end{pmatrix}
		\,.
		\label{eq:boundary_vector_for_eigen_function}
	\end{align}

\subsection{Classification of $\mathcal{CT}$ invariant boundary condition}

Since the parameter space of the general boundary condition \eqref{eq:BC_mode_function} is large and complicated to classify it,
we impose a symmetry to the system in order to obtain tractable boundary conditions.
 	
In this paper, we focus on the system which is invariant under the $\mathcal{CT}$ transformation
	\begin{align}
		&\Psi(t,y)\xrightarrow{\mathcal{CT}}\ \Psi'(t,y)=\sigma^3\Psi(-t,y)\,,
	\end{align}
where $\mathcal{C}$ indicates the charge conjugation and $\mathcal{T}$ denotes the time reversal.\footnote{
	In the representation \eqref{eq:gamma_rep},
	$\mathcal{C}$ and $\mathcal{T}$ transformations are given by
	\begin{align}
		\Psi(t,y)\xrightarrow{\mathcal{C}}\ \Psi'(t,y)=i\sigma^2\left(\Psi^\dagger(t,y)\sigma^1\right)^\mathrm{T}=\sigma^3\Psi^\ast(t,y)\,,
		&&
		\Psi(t,y)\xrightarrow{\mathcal{T}}\ \Psi'(t,y)=\Psi^\ast(-t,y)\,,
	\end{align}
up to phases.
}
The $\mathcal{CT}$ transformation for the mode functions is given as
	\begin{align}
		\Psi_{n}^{(i)}(y)\xrightarrow{\mathcal{CT}}\ \sigma^3\Psi_{n}^{(i)}(y)\,,\ \ \ \ \ 
		f_{n}^{(i)}(y)\xrightarrow{\mathcal{CT}}\ f_{n}^{(i)}(y)\,,\ \ \ \ \ 
		g_{n}^{(i)}(y)\xrightarrow{\mathcal{CT}}\ -g_{n}^{(i)}(y)\,.
	\end{align}	
Due to the $\mathcal{CT}$ symmetry and the anticommutativity of $\mathcal{H}_D$ and $\sigma^3$,
the negative energy eigenfunctions can be obtained from the positive ones as
	\begin{align} 
		\Psi_{-n}^{(i)}(y)=\sigma^3\Psi_{n}^{(i)}(y)
		=
		\begin{pmatrix}
		f_{n}^{(i)}(y)
		\\
		-g_{n}^{(i)}(y)
		\end{pmatrix}
		\ \ \ \ \ \ (n>0)\,.
	\end{align}
Therefore the $\mathcal{CT}$ symmetry ensures that the positive and negative energies are paired with each other.

In addition to the Dirac equation, the boundary condition should be also invariant under this transformation. Thus, we will focus on the $\mathcal{CT}$ invariant boundary condition
	\begin{align}
		&(\one_{2N}-U)\vec F_n^{(i)}=0\,,
		\label{eq:CT_BC_f}
		\\
		&(\one_{2N}+U)\vec G_n^{(i)}=0
		\label{eq:CT_BC_g}
	\end{align}
for all $n$ and $i$.
Although it seems that there are $4N$ conditions in \eqref{eq:CT_BC_f} and \eqref{eq:CT_BC_g}, we should totally obtain $2N$ conditions from \eqref{eq:BC_mode_function}. This imply that the eigenvalues of $U$ are $\pm1$ and therefore, $U$ is given by a Hermitian unitary matrix. 
This boundary condition corresponds to the one discussed in the previous paper~\cite{Fujimoto:2019fzb} and leads to the same structure of eigenfunctions. We can then classify the boundary condition and zero mode solutions in accordance with the previous paper.

We can then classify the matrix $U$ into $2N+1$ classes by the number of the eigenvalues $+1$ (or $-1$). We refer to the case with $K$ negative eigenvalues as the type $(2N-K,K)$ boundary condition $(K=0,1,\cdots,2N),$
and $U$ can be expressed as 
	\begin{align}
		\text{type}\,(2N-K,K):\qquad U&=V\left(\begin{array}{ccc:ccc}
		+1 &        &   0  &     &        &
		\\
		   & \ddots &      &     &   0    &
		\\
		 0 &        &  +1  &     &        &
		\\
		\hdashline 
		   &        &      &  -1 &        & 0
		\\
		   &   0    &      &     & \ddots &
		\\
		   &        &      &  0  &        & -1
		\\
		\end{array}
		\right)
		V^{\dagger}
		\label{eq:type2N-KKBC}\\
		&\hspace{1.1cm}{\underbrace{\hspace{2.2cm}}_{2N-K}\hspace{0.3cm}\underbrace{\hspace{2.2cm}}_{K}}\notag
	\end{align}
with a $2N\times2N$ unitary matrix $V$. 
\eqref{eq:type2N-KKBC} indicates that the parameter space of the type $(2N-K,K)$ boundary condition is given by the coset space $U(2N)/(U(2N-K)\times U(K))$. Since the continuous deformation of $V$ does not change the numbers of positive and negative eigenvalues in $U$, the different type of boundary conditions cannot be connected continuously.

Furthermore, we can express the $2N\times2N$ unitary matrix $V$ as 
	\begin{align}
		V=(\vec{u}_{1},\vec{u}_{2},\cdots,\vec{u}_{2N})\,,
		\label{eq:V_decomposition}
	\end{align}
where $\vec{u}_{r}\ (r=1,2,\cdots,2N)$ are $2N$-dimensional orthonormal complex vectors which satisfy $\vec{u}_{r}^{\dagger}\vec{u}_{r'}=\delta_{rr'}\ \ (r,r'=1,2,\cdots,2N)$.
Then, the matrix $U$ for the type $(2N-K,K)$ boundary condition is written by
	\begin{align}
		U=\sum_{p=1}^{2N-K}\vec{u}_{p}\vec{u}_{p}^{\dagger}-\sum_{q=2N-K+1}^{2N}\vec{u}_{q}\vec{u}_{q}^{\dagger}\,,
	\end{align}
and the boundary condition \eqref{eq:CT_BC_f} and \eqref{eq:CT_BC_g} are of the forms
	\begin{align}
		\vec{u}_{q}^{\dagger}\vec{F}_{n}^{(i)}&=0\qquad \text{for}\quad q=2N-K+1, 2N-K+2, \cdots, 2N\,,	
		\label{eq:BC-uF}\\
		\vec{u}_{p}^{\dagger}\vec{G}_{n}^{(i)}&=0\qquad \text{for}\quad p=1,\cdots,2N-K\,.\label{eq:BC-uG}
	\end{align}

\subsection{Zero mode solutions}

In this paper, we concentrate on the zero mode solutions $\Psi_{0,+}^{(i)}(y)=(f_0^{(i)}(y),0)^\mathrm{T}$ and  $\Psi_{0,-}^{(j)}(y)=(0,g_0^{(j)}(y))^\mathrm{T}$ which satisfy
	\begin{align}
		\mathcal{H}_D\Psi_{0,+}^{(i)}(y)=0\,,\ \ \ \ \ \ 
		\mathcal{H}_D\Psi_{0,-}^{(j)}(y)=0\,,
	\end{align}
or equivalently
	\begin{align}
		&(\partial_y+M)f_0^{(i)}(y)=0\,,
		\label{eq:Dirac_eq_f0}
		\\
		&(-\partial_y+M)g_0^{(j)}(y)=0\,.
		\label{eq:Dirac_eq_g0}
	\end{align}
	
The mode functions $f_0^{(i)}(y)$ and $g_0^{(j)}(y)$ on the rose graph can be discontinuous at the vertex and written as
	\begin{align}
		f^{(i)}_{0}(y)&=\sum^{N}_{a=1}\theta(y-L_{a-1})\theta(L_{a}-y)F^{(i)}_{a}C_{a}e^{-My}\,,\label{eq:f0}
		\\
		g^{(j)}_{0}(y)&=\sum^{N}_{a=1}\theta(y-L_{a-1})\theta(L_{a}-y)G^{(j)}_{a}C'_{a}e^{My}\label{eq:g0}\,,
	\end{align}
where $\theta(y)$ is the Heaviside step function and the constants $F_a^{(i)},\,G_a^{(j)}\in\mathbb{C}\ \ (a=1,\cdots,N)$ are determined by the boundary condition. Here we also introduced the constants $C_a$ and $C'_a$ which are given by
	\begin{align}
		C_a=\sqrt{\frac{1}{e^{-2M(L_{a-1}-\varepsilon)}+e^{-2M(L_{a}+\varepsilon)}}}
		\,,
		&&
		C'_a=\sqrt{\frac{1}{e^{2M(L_{a-1}-\varepsilon)}+e^{2M(L_{a}+\varepsilon)}}}\,,
		\label{eq:def_Ca_and_C'a}
	\end{align}
for later convenience.

We can find that if there are $m$ linearly independent solutions $f_0^{(i)}(y)\ (i=1,\cdots,m)$, there are also $m$ linearly independent $N$-dimensional complex vectors $\boldsymbol{F}^{(i)}\equiv (F^{(i)}_{1},F^{(i)}_{2},\cdots, F^{(i)}_{N})^{\rm T}\ (i=1,\cdots,m)$, and vice versa.
Similarly, the number of linearly independent solutions $g_0^{(j)}(y)\ (j=0,\cdots,m')$ corresponds to the number of linearly independent $N$-dimensional complex vectors $\boldsymbol{G}^{(j)}\equiv (G^{(j)}_{1},G^{(j)}_{2},\cdots, G^{(j)}_{N})^{\rm T}\ (j=0,\cdots,m')$.
\footnote{In this paper, we use the notation that a vector $\vec X$ with the vector symbol ``$\ \vec{\ }\ \ $'' indicates a $2N$-component vector and a vector $\boldsymbol X$ written by the bold symbol denotes an $N$-component vector.}

Then, let us discuss the independent zero modes under the type $(2N-K,K)$ boundary condition in terms of the vectors $\boldsymbol{F}^{(i)}$ and $\boldsymbol{G}^{(j)}$. For this purpose, it is convenient to introduce $2N$-dimensional vectors $\vec{\mathcal{F}}_{a}$ and $\vec{\mathcal{G}}_{a}\ (a=1,\cdots,N)$ 
	\begin{align}
		\vec{\mathcal{F}}_{a}&\equiv C_{a}(\underbrace{0,\cdots,0}_{{2(a-1)}}, e^{-M(L_{a-1}+\varepsilon)},e^{-M(L_{a}-\varepsilon)},0,\cdots,0)^{\rm T},
		\label{eq:vec_F}
		\\
		\vec{\mathcal{G}}_{a}&\equiv C'_{a}(\underbrace{0,\cdots,0}_{{2(a-1)}}, e^{M(L_{a-1}+\varepsilon)},-e^{M(L_{a}-\varepsilon)},0,\cdots,0)^{\rm T}.
		\label{eq:vec_G}
	\end{align}
The constants $C_a$ and $C_a'$ are the same as \eqref{eq:def_Ca_and_C'a}. These vectors are orthonormal
	\begin{align}
		\vec{\mathcal{F}}_{a}^{\dagger}\vec{\mathcal{F}}_{b}=\vec{\mathcal{G}}_{a}^{\dagger}\vec{\mathcal{G}}_{b}=\delta_{ab}
		\,,
		&&
		\vec{\mathcal{F}}_{a}^{\dagger}\vec{\mathcal{G}}_{b}=\vec{\mathcal{G}}_{a}^{\dagger}\vec{\mathcal{F}}_{b}=0
		\qquad (a,b=1,2,\cdots,N)\,,\label{eq:FFGG-normalization}
	\end{align}
and form a complete set in the $2N$-dimensional complex vector space.

By using the vectors \eqref{eq:vec_F} and \eqref{eq:vec_G},  we can express the boundary vectors $\vec{F}^{(i)}_{0},\ \vec{G}^{(j)}_{0}$ in Eq.~\eqref{eq:boundary_vector_for_eigen_function} for $n=0$ and $2N$-dimensional complex vectors $u_p\ (p=1,\cdots,2N)$ in \eqref{eq:V_decomposition} as follows:
	\begin{align}
		\vec{F}^{(i)}_{0}&=\sum^{N}_{a=1}F_{a}^{(i)}\vec{\mathcal{F}}_{a},\label{eq:F-decomposition}\\
		\vec{G}^{(j)}_{0}&=\sum^{N}_{a=1}G_{a}^{(j)}\vec{\mathcal{G}}_{a},\label{eq:-decomposition}
		\\
		\vec{u}_{r}&=\sum^{N}_{a=1}\alpha_{r,{a}}\vec{\mathcal{F}}_{a}+\sum^{N}_{a=1}\beta_{r,a}\vec{\mathcal{G}}_{a}\label{eq:2Nvector-decomposition}
		\qquad(r=1,2,\cdots,2N)\,,
	\end{align}
where $\alpha_{r,a}$ and $\beta_{r,a}$ are complex constants and satisfy
	\begin{align}
		\sum_{a=1}^N \left(\alpha_{r,{a}}^\ast\alpha_{r',{a}}+\beta_{r,{a}}^\ast\beta_{r',{a}}\right)=\delta_{rr'}\qquad(r,r'=1,2,\cdots,2N)
		\label{eq:alpha_beta_condition}
	\end{align}
from the orthonormal relations for $\vec{u}_{r}$.

The boundary condition \eqref{eq:BC-uF} and \eqref{eq:BC-uG} for $n=0$ can be rewritten as
	\begin{align}
		\boldsymbol{\alpha}^{\dagger}_{q}\,{\cdot}\,\boldsymbol{F}^{(i)}&=0\,,\qquad (q=2N-K+1,\cdots,2N)\,,\label{eq:alphaF-condition}
		\\
		\boldsymbol{\beta}^{\dagger}_{p}\,{\cdot}\,\boldsymbol{G}^{(j)}&=0\,,\qquad (p=1,\cdots,2N-K)\,,\label{eq:betaG-condition}
	\end{align}
where $\boldsymbol\alpha_{q}\equiv (\alpha_{q,1},\alpha_{q,2},\cdots,\alpha_{q,N})^{{\rm T}}$,
\ $\boldsymbol\beta_{p}\equiv (\beta_{p,1},\beta_{p,2},\cdots,\beta_{p,N})^{{\rm T}}$ and it turns out that $\boldsymbol{F}^{(i)}\ (\boldsymbol{G}^{(j)})$ are given by vectors which are orthogonal to $\boldsymbol\alpha_{q}\ (\boldsymbol{\beta}_p)$ for $q=2N-K+1,\cdots,2N\ (p=1,\cdots,2N-K)$.

Let us suppose that the number of the linearly independent vectors for $\boldsymbol\alpha_{q}\,\ (q=2N-K+1,\cdots,2N)$ is $l$. Since the $2N$-component vectors $\vec{u}_{p}\ (p=1,\cdots,2N-K)$ and $\vec{u}_{q}\ (q=2N-K+1,\cdots,2N)$ should be linearly independent with each other,
the number of the linearly independent vectors for $\boldsymbol\beta_{q}\ (q=2N-K+1,\cdots,2N)$, $\boldsymbol\alpha_{p}$ and $\boldsymbol\beta_{p}\ (p=1,\cdots,2N-K)$ are given as $K-l$, $N-l$ and $N-K+l$, respectively. Therefore the range of $l$ is restricted to $0\leq l \leq K$ for the case of $K=0,\cdots,N$ and $K-N\leq l \leq N$ for the case of $K=N,\cdots,2N$.

If the number of the linearly independent vectors for $\boldsymbol\alpha_{q}\ (q=2N-K+1,\cdots,2N)$ and $\boldsymbol\beta_{p}\ (p=1,\cdots,2N-K)$ are $l$ and $N-K+l$ respectively, we can find that there exist $N-l$ linearly independent solutions for $\boldsymbol{F}^{(i)}\ (i=1,\cdots,N-l)$ and $K-l$ linearly independent solutions for $\boldsymbol{G}^{(j)}\ (j=1,\cdots,K-l)$ from Eqs.~\eqref{eq:alphaF-condition} and \eqref{eq:betaG-condition}.
These also imply that there are $N-l$ linearly independent boundary vectors for $\vec{F}^{(i)}_0\ (i=1,\cdots,N-l)$ and $K-l$ linearly independent boundary vectors for $\vec{G}^{(j)}_0\ (j=1,\cdots,K-l)$.
Therefore, for the type $(2N-K,K)$ boundary condition, we can conclude that 
the degeneracy of zero mode $f^{(i)}_0(y)$ is given by $N_{f_0}=N-l$, and
the one of zero mode $g^{(i)}_0(y)$ is given by $N_{g_0}=K-l$.
The degeneracies $N_{f_0}$ and $N_{g_0}$ for each boundary condition are described in Table \ref{tab:number_zero_mode}.

It should be noted that the difference $N_{f_{0}}-N_{g_{0}}(=N-K)$ is independent of $l$ and invariant under continuous deformations of the boundary condition (although $l$ can be changed by those deformations).
This is because that the structure of the supersymmetric quantum mechanics is hidden in this $\mathcal{CT}$ invariant system as well as in Ref.~\cite{Fujimoto:2019fzb}.
In this system, the ``Hamiltonian'' $H$\,,\ the supercharge $Q$ and the ``fermion number'' operator $(-1)^F$ in the supersymmetric quantum mechanics can be given as $H=Q^2,\ Q=\mathcal{H}_D,\ (-1)^F=\sigma^3$. They are well-defined and Hermitian in the $\mathcal{CT}$ invariant boundary condition.
Then, we can introduce the Witten index $\Delta_W$ known as the topological quantity, which is defined by the difference of the numbers of the solutions with the eigenvalues $Q=0$ and $(-1)^F=\pm1$. Therefore, $\Delta_W=N_{f_{0}}-N_{g_{0}}$ is invariant under continuous deformations of the boundary condition, as it should be.

In the next section, we will see these degenerate zero modes lead to non-Abelian Berry's connections.
\\

\begin{table}[h]
\centering	
	\label{table:2N-kBC}
	{
		\begin{tabular}{c|c|c|c|c}
			\hline
			$K$ &$l$&$N_{f_{0}}$&$N_{g_{0}}$&$N_{f_{0}}-N_{g_{0}}$\\
			\hline 
			&$0$&$N$&$K$&$N-K$\\
			&$1$&$N-1$&$K-1$&$N-K$\\
			$0\leq K\leq N$&\vdots&\vdots&\vdots&\vdots\\
			&$K-1$&$N-K+1$&$1$&$N-K$\\
			&$K$&$N-K$&$0$&$N-K$\\
			\hline
			&$K-N$&$2N-K$&$N$&$N-K$\\
			&$K-N+1$&\ $2N-K-1$\ &\ $N-1$\ &$N-K$\\
			$N\leq K\leq 2N$&\vdots&\vdots&\vdots&\vdots\\
			&$N-1$&$1$&$K-N+1$&$N-K$\\
			&$N$&$0$&$K-N$&$N-K$\\
			\hline
		\end{tabular}
	}
    \captionsetup{width=.85\linewidth}
	\caption{The number of the zero mode solutions of $f^{(i)}_{0}(y)$ and $g^{(j)}_{0}(y)$ for the {type\,($2N-K,K$)} boundary condition. $l$ denotes the maximal number of the linearly independent vectors $\boldsymbol\alpha_{q}\,(q=2N-K+1,\cdots,2N)$ in Eq.~(\ref{eq:2Nvector-decomposition}). $N_{f_{0}}$ ($N_{g_{0}}$) is the number of the zero mode solutions of $f^{(i)}_{0}(y)$ ($g^{{(j)}}_{0}(y)$).
	We can find that difference $N_{f_{0}}-N_{g_{0}}(=N-K)$ is independent on $l$, though both of $N_{f_0}$ and $N_{g_0}$ depend on $l$.}
	\label{tab:number_zero_mode}
\end{table}

\section{Non-Abelian Berry's connection on quantum graph}
\label{sec:berrys-connection}
In this section, we discuss non-Abelian Berry's connections for the degenerate zero modes in the parameter space of the boundary conditions and clarify how instantons appear as the Berry's connections by using the method of the ADHM construction.

\subsection{Berry's connection for zero modes}
\label{sbsec:berrys-connection_for_zero_modes}

Here, we consider the situation that the parameters
of the boundary condition $\boldsymbol\alpha_{q}\ (q=2N-K+1,\cdots,2N)$ and $\boldsymbol\beta_{p}\ (p=1,\cdots,2N-K)$ in the type $(2N-K,K)$ boundary condition are time-dependent, and vary adiabatically along closed paths ({\it i.e.} 
$\boldsymbol\alpha_{q}(t=0)=\boldsymbol\alpha_{q}(t=T)$ and  $\boldsymbol\beta_{p}(t=0)=\boldsymbol\beta_{p}(t=T)$)
without any change of the numbers of their linearly independent vectors $l$ and $N-K+l$ respectively.

Then, let us solve the Dirac equations
	\begin{align}
		i\frac{\partial}{\partial t}\Psi_{0,+}^{(i)}(t,y)&=\mathcal{H}_D\Psi_{0,+}^{(i)}(t,y)\,,
		\\
		i\frac{\partial}{\partial t}\Psi_{0,-}^{(j)}(t,y)&=\mathcal{H}_D\Psi_{0,-}^{(j)}(t,y)
	\end{align}
with the initial conditions $\Psi_{0,+}^{(i)}(0,y)=\Psi_{0,+}^{(i)}(y\,;\alpha_{q}(0))$ and $\Psi_{0,-}^{(j)}(0,y)=\Psi_{0,-}^{(j)}(y\,;\beta_{p}(0))$.
Although the final states $\Psi_{0,+}^{(i)}(T,y)$ and $\Psi_{0,-}^{(j)}(T,y)$ continue to be eigenstates with $\mathcal{H}_D=0$ under this time-evolution due to the adiabatic theorem, they are given as linear combinations of the degenerate initial states at $t=0$~\cite{Wilczek:1984dh}:
	\begin{align}
		\Psi_{0,+}^{(i)}(T,y)&=\sum_{i'=1}^{N-l}\Psi_{0,+}^{(i')}(y\,;\boldsymbol\alpha_{q}(0))\,(\Gamma_f[C_f])_{i'i}\ \ \ \ (i=1,\cdots,N-l)\,,
		\\
		\Psi_{0,-}^{(j)}(T,y)&=\sum_{j'=1}^{K-l}\Psi_{0,-}^{(j')}(y\,;\boldsymbol\beta_{p}(0))\,(\Gamma_g[C_g])_{j'j}\ \ \ \ (j=1,\cdots,K-l)\,,
	\end{align}
where $C_{f}$ and $C_g$ denote closed paths on the parameter spaces of $\boldsymbol\alpha_{q}\ (q=2N-K+1,\cdots,2N)$ and $\boldsymbol\beta_{p}\ (p=1,\cdots,2N-K)$, respectively.\footnote{We can consider time-evolutions of  $\Psi_{0,+}^{(i)}$ and  $\Psi_{0,-}^{(i)}$ individually. This is because the boundary condition \eqref{eq:alphaF-condition} and \eqref{eq:betaG-condition} are separated for the zero modes $\Psi_{0,\pm}^{(i)}$ and independent with each other, provided that the number of their linearly independent vectors $\boldsymbol\alpha_{q}\,\ (q=2N-K+1,\cdots,2N)$ does not change.} The matrices $\Gamma_{f}[C_f]$ and $\Gamma_{g}[C_g]$ are called the non-Abelian Berry's phase and given as the path-ordered exponential
	\begin{align}
		\Gamma_f[C_f]&=\mathcal{P}\exp\left(-\oint_{C_f} A^{(f)}\right)\,,
		\\
		\Gamma_g[C_g]&=\mathcal{P}\exp\left(-\oint_{C_g} A^{(g)}\right)\,.
	\end{align}
$A^{(f)}$ and $A^{(g)}$ are anti-Hermitian matrices of 1-forms defined by
	\begin{align}
		(A^{(f)})_{i'i}
		&\equiv \int dy\,\left(\Psi_{0,+}^{(i')}(y,\boldsymbol\alpha_{q})\right)^\dagger d\Psi_{0,+}^{(i)}(y,\boldsymbol\alpha_{q})
		\ \ \ \ (i,i'=1,\cdots,N-l)\,,
		\label{eq:Berry_connection_f}
		\\
		(A^{(g)})_{j'j}&\equiv \int dy\,\left(\Psi_{0,-}^{(j')}(y,\boldsymbol\beta_{p})\right)^\dagger d\Psi_{0,-}^{(j)}(y,\boldsymbol\beta_{p})\ \ \ \ (j,j'=1,\cdots,K-l)
		\label{eq:Berry_connection_g}
	\end{align}
with  the exterior derivative $d$ for the parameter space of the boundary condition.
These 1-forms are known as the non-Abelian Berry's connections. 
Under the following unitary transformations for the  zero modes
	\begin{align}
		\Psi_{0,+}^{(i)}&\to\sum_{i'=1}^{N-l}\Psi_{0,+}^{(i')}(U_{f})_{i'i}\ \ \ \ (i=1,\cdots,N-l)\,,
		\\
		\Psi_{0,-}^{(j)}&\to\sum_{j'=1}^{K-l}\Psi_{0,-}^{(j')}(U_{g})_{j'j}\ \ \ \ (j=1,\cdots,K-l)\,,
	\end{align}
$A^{(f)}$, $A^{(g)}$ and $\Gamma_f[C_f]$, $\Gamma_g[C_g]$ transform as
	\begin{align}
		A^{(f)}&\to U_f^{-1}A^{(f)}U_f+U_f^{-1}dU_f\,,&\Gamma_{f}[C_{f}]&\to U_f^{-1}\Gamma_f[C_f]U_f\,,
		\\
		A^{(g)}&\to U_g^{-1}A^{(g)}U_g+U_g^{-1}dU_g\,,
		&
		\Gamma_{g}[C_{g}]&\to U_g^{-1}\Gamma_g[C_g]U_g\,.
	\end{align}
Therefore $A^{(f)}$ and $A^{(g)}$ are just like gauge potentials.

Since the zero modes depend on the boundary condition through only the coefficients $F^{(i)}_{a}$ and $G^{(j)}_{a}\ (a=1,\cdots,N)$, we can rewrite \eqref{eq:Berry_connection_f} and \eqref{eq:Berry_connection_g} by using the expressions \eqref{eq:f0} and \eqref{eq:g0} as follows:
	\begin{align}
		A^{(f)}=\tilde F^\dagger(\boldsymbol\alpha_{q})\,d\tilde F(\boldsymbol\alpha_{q})\,,
		\label{eq:Berry's_connection_f}
		\\
		A^{(g)}=\tilde G^\dagger(\boldsymbol\beta_{p})\,d\tilde G(\boldsymbol\beta_{p})\,,
		\label{eq:Berry's_connection_g}
	\end{align}
where $\tilde F(\boldsymbol\alpha_{q})$ and $\tilde G(\boldsymbol\beta_{p})$ are the matrices defined by the vectors $\boldsymbol{F}^{(i)}=(F^{(i)}_{1},F^{(i)}_{2},\cdots, F^{(i)}_{N})^{\rm T}$ and  $\boldsymbol{G}^{(j)}=(G^{(j)}_{1},G^{(j)}_{2},\cdots, G^{(j)}_{N})^{\rm T}$ with the $N\times N$ matrix $\mathcal{C}$
	\begin{align}
		\tilde F&=\mathcal{C}F\,,\ \ \ \ \ \ 
		F=
		\begin{pmatrix}
		 \boldsymbol{F}^{(1)}\ \ \boldsymbol{F}^{(2)}\ \ \cdots\ \ \boldsymbol{F}^{(N-l)}
		\end{pmatrix}\,,
		\\
		\tilde G&=\mathcal{C}G\,,\ \ \ \ \ \  
		G=\begin{pmatrix}
		\boldsymbol{G}^{(1)}\ \ \boldsymbol{G}^{(2)}\ \ \cdots\ \ \boldsymbol{G}^{(K-l)}
		\end{pmatrix}\,,
		\\
		\mathcal{C}&=\begin{pmatrix}
		\sqrt{\frac{\tanh M(L_1-L_{0})}{2M}} & & 0
		\\
		& \ddots &
		\\
		0 & & \sqrt{\frac{\tanh M(L_N-L_{N-1})}{2M}}
		\end{pmatrix}\,,
	\end{align}
and should satisfy the relations
	\begin{align}
		\tilde F^\dagger \tilde F&=\one_{N-l}\,,
		\label{eq:normalization_f}
		\\
		\tilde G^\dagger \tilde G&=\one_{K-l}
		\label{eq:normalization_g}
	\end{align}
due to the normalization conditions of the zero mode functions.

\subsection{Instantons in Berry's connections}
\label{sbsec:instantons_in_the_berry's_connections}

To discuss the instantons in the Berry's connections, we consider the case that all the edges in the rose graph have the same length, {\it i.e.}, $L_1-L_0=\cdots=L_{N}-L_{N-1}=L$.  
In this case, the boundary condition \eqref{eq:alphaF-condition} and \eqref{eq:betaG-condition} are equivalent to
	\begin{align}
		\alpha^\dagger_{[l]\times[N]} \tilde F_{[N]\times[N-l]}=0\,,
		\label{eq:BC_f}
		\\
		\beta^\dagger_{[N-K+l]\times[N]} \tilde G_{[N]\times[K-l]}=0\,,
		\label{eq:BC_g}
	\end{align}
where $\alpha$ is an $N\times l$ matrix whose column vectors are given by linear combinations of the vectors $\boldsymbol\alpha_{q}\ (q=2N-K+1,\cdots,2N)$ and independent with each other. $\beta$ is also an $N\times (N-K+l)$ matrix whose column vectors are independent with each other and given by linear combinations of the vectors $\boldsymbol\beta_{p}\ (p=1,\cdots,2N-K)$.

Then, a crucial observation is the following correspondence between the Berry's connections on the quantum graph with $\alpha,\,\tilde F,\,A^{(f)}$ and the ADHM construction of the  instantons with $\Delta(x),\,v,\,A$ as discussed in Section~\ref{sec:instantons-and-the-adhm-construction}:
 	\begin{align}
 		&\text{Boundary condition} &&&& \text{Zero mode equation}
 		\notag\\
		&\alpha^\dagger_{[l]\times[N]} \tilde F_{[N]\times[N-l]}=0 && \Longleftrightarrow && \Delta^\dagger(x)_{[2k]\times[n+2k]}v_{[n+2k]\times[n]}=0\,,
		\notag\\
		\notag\\
		&\text{Normalization condition} &&&& \text{Normalization condition}\notag
		\notag\\
		&\tilde F^\dagger\tilde F=\one_{[N-l]\times[N-l]} && \Longleftrightarrow &&
		v^\dagger v=\one_{[n]\times[n]}\,,
		\notag\\
		\notag\\
	    &\text{Berry's connection} &&&& \text{Instanton}\notag
		\notag\\
		&A^{(f)}=\tilde F^\dagger d\tilde F
		&& \Longleftrightarrow &&
		A=v^\dagger dv\,.
	\end{align}
The case for $A^{(g)}$ has also the same correspondence. Here, the exterior derivative in the Berry's connection acts on the parameter space of the boundary condition while the one in the instanton acts on the coordinates of $\mathbb{R}^4$.

Therefore, in the case of even $l$, we find that the Berry's connection $A^{(f)}$ is given as the ${\it SU}(N-l)$ instanton with the topological charge $|Q|=l/2$ if we parametrize the matrix $\alpha$ as 
	\begin{align}
		\alpha_{[N]\times[l]}&=\Delta(x)_{[N]\times[l]}\,,
		\label{eq:ADHM_alpha}
	\end{align}
and only the parameters $x^\mu$ vary adiabatically depending on the time while the others are fixed. Here $\Delta(x)_{[N]\times[l]}$ is the matrix of the ADHM construction whose size is $N\times l$.
Similarly, in the case that $N-K+l$ is even, the Berry's connection $A^{(g)}$ is given as the ${\it SU}(K-l)$ instanton with the topological charge $|Q|=(N-K+l)/2$
if we parametrize the matrix $\beta$ as 
	\begin{align}
		\beta_{[N]\times[N-K+l]}&=\Delta(x)_{[N]\times[N-K+l]}\,,
		\label{eq:ADHM_beta}
	\end{align}
and only the parameters $x^\mu$ vary adiabatically depending on the time.\footnote{It should be noted that there are degrees of freedom by the redefinitions $\alpha\to\alpha R^{(\alpha)}\ \ (R^{(\alpha)}\in GL(l;\mathbb{C}))$ and $\beta\to\beta R^{(\beta)}\ \ (R^{(\beta)}\in GL(N-K+l;\mathbb{C}))$ which does not affect the boundary condition \eqref{eq:BC_f}, \eqref{eq:BC_g} and zero modes. }
The gauge groups and the topological charges of the instantons in the case of $N=4,5,6$ are summarized in Table~\ref{table:gauge_group}.
There does not exist instantons in the cases of $N=2,3$.
\begin{table}[h]
	\centering	
	{\small
		\begin{tabular}{ccc|cc|cc}
			\hline
			& & & gauge group of $A^{(f)}$ & topological charge & gauge group of $A^{(g)}$ & topological charge
			\\
			\hline
			$N$ & $K$  & $l$  &  ${\it SU}(N-l)$  &  $|Q|=l/2$   &   $\it SU(K-l)$&  $|Q|=(N-K+l)/2$
			\\
			\hline
			&   2  &  0  &     --      &    --   &  {\it SU}(2) & 1
			\\
			&   2  &  2  &    {\it SU}(2) &   1     &  --          & -- 
			\\
			&   3  &  1  &    --          &   --    & {\it SU}(2)  & 1
			\\
			&   3  &  2  &    {\it SU}(2) &   1     &  --          & --  
			\\
			4   &   4  &  2  &    {\it SU}(2) &   1     & {\it SU}(2)  & 1
			\\
			&   5  &  2  &    {\it SU}(2) &   1     & --           & --
			\\
			&   5  &  3  &    --          &   --    & {\it SU}(2)  & 1
			\\
			&   6  &  2  &    {\it SU}(2) &   1     & --           & --
			\\
			&   6  &  4  &    --          &   --    & {\it SU}(2)  & 1
			\\
			\hline
			&   2  &  2  &    {\it SU}(3) &   1     & --           & --
			\\ 
			&   3  &  0  &    --          &   --    & {\it SU}(3)  & 1
			\\
			&   3  &  2  &    {\it SU}(3) &   1     &  --          & --
			\\
			&   4  &  1  &    --          &   --    & {\it SU}(3)  & 1
			\\
			&   4  &  2  &    {\it SU}(3) &   1     &  --          & --
			\\
			5  &   5  &  2  &    {\it SU}(3) &   1     & {\it SU}(3)  & 1
			\\
			&   6  &  2  &    {\it SU}(3) &   1     & --           &  --
			\\
			&   6  &  3  &    --          &   --    & {\it SU}(3)  & 1
			\\
			&   7  &  2  &    {\it SU}(3) &   1     & --             & --
			\\
			&   7  &  4  &    --          &   --    & {\it SU}(3)  & 1
			\\
			&   8  &  5  &    --          &  --     & {\it SU}(3)  & 1
			\\
			\hline
			     &   2  &  0  &    --          &  --     & {\it SU}(2)  & 2
			 \\
			     &   2  &  2  &    {\it SU}(4) &   1     & --           & --
			 \\
			     &   3  &  1  &    --          & --      & {\it SU}(2)  & 2
			 \\
			     &   3  &  2  &    {\it SU}(4) &   1     & --           & --
			 \\
			     &   4  &  0  &    --          &  --     & {\it SU}(4)  & 1
			 \\
			     &   4  &  2  &    {\it SU}(4) &   1     & {\it SU}(2)  & 2
			 \\
			     &   4  &  4  &    {\it SU}(2) &   2     & --           & --
			 \\
			     &   5  &  1  &    --          &  --     & {\it SU}(4)  & 1
			 \\  
			     &   5  &  2  &   {\it SU}(4)  &   1     &  --          & --
			 \\  
			     &   5  &  3  &   --           &  --     & {\it SU}(2)  & 2
			 \\
			     &   5  &  4  &   {\it SU}(2)  &   2     & --           & --
			 \\
			     &   6  &  2  &   {\it SU}(4)  &   1     & {\it SU}(4)  & 1
			 \\
			  6  &   6  &  4  &   {\it SU}(2)  &   2     & {\it SU}(2)  & 2
			 \\
			     &   7  &  2  &   {\it SU}(4)  &   1     & --           & --
			 \\
			     &   7  &  3  &   --           & --      & {\it SU}(4)  & 1
			 \\
			     &   7  &  4  &   {\it SU}(2)  &   2     & --           & --
			 \\
			     &   7  &  5  &   --           &  --     & {\it SU}(2)  & 2
			 \\
			     &   8  &  2  &   {\it SU}(4)  &   1     & --           & --
			 \\
			     &   8  &  4  &   {\it SU}(2)  &   2     & {\it SU}(4)  & 1
			 \\
			     &   8  &  6  &   --           &  --     & {\it SU}(2)  & 2
			 \\
			     &   9  &  4  &   {\it SU}(2)  &   2     &  --          & --
			 \\
			     &   9  &  5  &    --          & --        & {\it SU}(4)  & 1
			 \\
			     &  10  &  4  &   {\it SU}(2)  &   2     &  --          & --
			 \\
			     &  10  &  6  &    --          &   --    & {\it SU}(4)  & 1
			 \\    \hline
		\end{tabular}
	}
	\captionsetup{width=.85\linewidth}
	\caption{The gauge groups of instantons and topological charges in the case of $N=4,5,6$\,. The lines indicate that the instanton does not exist. The cases without the instantons for both $A^{(f)}$ and $A^{(g)}$ are not listed in this table.}
	\label{table:gauge_group}
\end{table}

Then, let us see the concrete examples for $(N,K,l)=(4,2,2)$ and $(6,4,4)$ below.
\\
\\
$\bullet$ {\bf The case of $(N,K,l)=(4,2,2)$}

In this case, there are two independent zero modes for $f^{(i)}_{0}$ and  
the ${\it SU}(2)$ instantons with the topological charge $Q=\pm1$ appear as the Berry's connections $A^{(f)}$. The vectors $\boldsymbol\alpha_{q}\ (q=7,8)$ are linearly independent with each other since $l=2$. Here we take them as
	\begin{align}
		\boldsymbol\alpha_7&=
		\frac{1}{\sqrt{\rho^2+(x-b)^2}}\begin{pmatrix}
			\rho
			\\
			0
			\\
			x^4-b^4-i(x^3-b^3)
			\\
			x^2-b^2-i(x^1-b^1)
		\end{pmatrix}
		\,,
		\ \ \ \ 
		\boldsymbol\alpha_8=
		\frac{1}{\sqrt{\rho^2+(x-b)^2}}\begin{pmatrix}
		0
		\\
		\rho
		\\
		-(x^2-b^2)-i(x^1-b^1)
		\\
		x^4-b^4+i(x^3-b^3)
		\end{pmatrix}\,,
	\end{align}
where $b^\mu,\,\rho\in\mathbb{R}$, and also take the matrix $\alpha$ as
	\begin{align}
		\alpha_{[4]\times[2]}&=(\sqrt{\rho^2+(x-b)^2}\boldsymbol\alpha_7\ \ \sqrt{\rho^2+(x-b)^2}\boldsymbol\alpha_8)
		\notag\\
		&=
		\begin{pmatrix}
		\rho\ \ \ \  0 
		\\
		0\ \ \ \ \rho \vspace{0.2cm}
		\\
		e_\mu(x^\mu-b^\mu)
		\vspace{0.2cm}
		\end{pmatrix}
		\,.
	\end{align}
This $\alpha$ corresponds to the canonical form of the ADHM data \eqref{eq:canonical_form}. Then, the matrix $\tilde F$ which satisfies Eq.~\eqref{eq:BC_f} is given by
	\begin{align}
		\tilde F_{[4]\times[2]}=\frac{1}{\sqrt{\rho^2+(x-b)^2}}
		\begin{pmatrix}
		\vspace{-0.2cm}
		\\
		\vspace{0.2cm}
		e_\mu^\dagger(x^\mu-b^\mu)
		\\
		\hspace{-0.2cm}-\rho\hspace{0.2cm}\ \ \ \  0
		\\
		0\ \ \ \ -\rho 
		\end{pmatrix}\,.
	\end{align}
Therefore, when we consider the situation that the parameter $x^\mu$ vary adiabatically, we obtain the Berry's connection
	\begin{align}
		A^{(f)}(x)=-i\eta^{(-)}_{\mu\nu}\frac{(x-b)^\nu}{\rho^2+(x-b)^2}dx^\mu\,,
	\end{align}
where $(x-b)^2=(x-b)^\mu(x-b)^\mu$ and $\eta^{(-)}_{\mu\nu}\equiv-\frac{i}{2}(e_\mu e_\nu^\dagger-e_\nu  e_\mu^\dagger)$ is called the 't Hooft $\eta$ symbol. This connection is well known as the Belavin-Polyakov-Schwartz-Tyupkin (BPST) instanton~\cite{Belavin:1975fg} and has the topological charge $-1$. Here, $b^\mu $ and $\rho$ correspond to the position and size of the instanton, respectively. 
\\
\\
$\bullet$ {\bf The case of $(N,K,l)=(6,4,4)$}

In this case, there are two independent zero modes for $f^{(i)}_{0}$ and the Berry's connection can be given as the ${\it SU}(2)$ instantons with the topological charge $Q=\pm2$. The vectors $\boldsymbol\alpha_{q}\ (q=9,10,11,12)$ are linearly independent with each other. Here we focus on the case that they are parametrized by
	\begin{alignat}{2}
		\boldsymbol\alpha_9&=
		\frac{1}{\sqrt{\rho_1^2+(x-b_1)^2}}\begin{pmatrix}
		\rho_1
		\\
		0
		\\
		x^4-b_1^4-i(x^3-b_1^3)
		\\
		0
		\\
		x^2-b_1^2-i(x^1-b_1^1)
		\\
		0
		\end{pmatrix}
		\,,
		& 
		\boldsymbol\alpha_{10}&=
		\frac{1}{\sqrt{\rho_2^2+(x-b_2)^2}}\begin{pmatrix}
		\rho_2
		\\
		0
		\\
		0
		\\
		x^4-b_2^4-i(x^3-b_2^3)
		\\
		0
		\\
		x^2-b_2^2-i(x^1-b_2^1)
		\end{pmatrix}
		\,,
		\notag\\
		\boldsymbol\alpha_{11}&=
		\frac{1}{\sqrt{\rho_1^2+(x-b_1)^2}}\begin{pmatrix}
		0
		\\
		\rho_1
		\\
		-(x^2-b_1^2)-i(x^1-b_1^1)
		\\
		0
		\\
		x^4-b_1^4+i(x^3-b_1^3)
		\\
		0
		\end{pmatrix}
		\,,\ \ 
		&
		\boldsymbol\alpha_{12}&=
		\frac{1}{\sqrt{\rho_2^2+(x-b_2)^2}}\begin{pmatrix}
		0
		\\
		\rho_2
		\\
		0
		\\
		-(x^2-b_2^2)-i(x^1-b_2^1)
		\\
		0
		\\
		x^4-b_2^4+i(x^3-b_2^3)
		\end{pmatrix}
		\,,
		\notag\\
		&&&(\rho_i\,,\ b^\mu_i\in\mathbb{R}\ \ (i=1,2))\,.
	\end{alignat}
Then, we take the matrix $\alpha$ as
	\begin{align}
		\alpha_{[6]\times[4]}&=(\boldsymbol\alpha_9\ \ \boldsymbol\alpha_{10}\ \ \boldsymbol\alpha_{11}\ \ \boldsymbol\alpha_{12})\cdot
		\left[\one_2\otimes\begin{pmatrix}
			\sqrt{\rho_1^2+(x-b_1)^2}  & 0   
			\\
			0 & \sqrt{\rho_2^2+(x-b_2)^2} 
		\end{pmatrix}\right]
		\notag\\
		&=
		\begin{pmatrix}
		\rho_1\ \ \ \ \rho_2 \ \ \ \ 0 \ \ \ \ 0
		\\
		\ \ 0\ \ \ \ \ \ 0\ \ \ \ \ \rho_1\ \ \ \ \rho_2 \vspace{0.2cm}
		\\
		e_\mu\otimes (x^\mu-T^\mu)
		\vspace{0.2cm}
		\end{pmatrix}
		\,,
	\end{align}
where $T^\mu_{[2]\times[2]}={\rm diag}(b^\mu_1,b^\mu_2)$, and we obtain
	\begin{align}
		\tilde F_{[6]\times[2]}=\frac{1}{\sqrt{1+\sum_{i=1}^2\frac{\rho_i^2}{(x-b_i)^2}}}
		\begin{pmatrix}
		\one_2
		\\
		-e_\mu\otimes\left(\rho_1\frac{(x-b_1)^\mu}{(x-b_1)^2}\,,\ \rho_2\frac{(x-b_2)^\mu}{(x-b_2)^2}\right)^\dagger
		\end{pmatrix}\,.
	\end{align}
Therefore, when we consider the situation that the parameter $x^\mu$ vary adiabatically, we obtain the Berry's connection as
	\begin{align}
		A^{(f)}(x)=-i\eta^{(+)}_{\mu\nu}\frac{\sum_{i=1}^2\rho_i^2\frac{(x-b_i)^\nu}{(x-b_i)^4}}{1+\sum_{i=1}^2\frac{\rho_i^2}{(x-b_i)^2}}dx^\mu
	\end{align}
with the 't Hooft $\eta$ symbol $\eta^{(+)}_{\mu\nu}\equiv-\frac{i}{2}(e_\mu^\dagger e_\nu- e_\nu^\dagger e_\mu)$. This connection is the 't Hooft instanton with the topological charges $Q=-2$.

\section{Conclusion and Discussion}
\label{sec:conclusion-and-discussion}

In this paper, we have considered the Dirac zero modes on the quantum graph and studied the non-Abelian Berry's connections in the parameter space of the $\mathcal{CT}$ invariant boundary conditions.
Then, we revealed the parameter space which gives the ${\it SU}(n)$ instantons as the Berry's connections due to the structure of the ADHM construction.
Although we have considered the 1+1d system, the same results can be obtained in the case of other dimensions with a quantum graph when we restrict the boundary condition to the type of \eqref{eq:CT_BC_f} and \eqref{eq:CT_BC_g}.
Furthermore, the constructions for the $O(n)$ and ${\it Sp}(n)$ instantons are known and will be applied to this system.
The higher dimensional generalizations of ADHM construction are also discussed in~\cite{Nakamula:2016srw,Takesue:2017gjw,Loginov:2020evw} and higher dimensional instantons would appear as the Berry's connections.

Since we have restricted the parameter space, other topological configurations such as  monopoles, vortices and so on may appear in the other parameter spaces. Especially, the method for constructing general Bogomolny--Prasad--Sommerfield (BPS) monopole solutions is known as the Nahm construction and applied to the Berry's connection of ${\it SU}(2)$ monopoles in various models, e.g. supersymmetric quantum mechanics~\cite{Wong:2015cnt}, Weyl semimetal~\cite{Hashimoto:2016dtm} and furthermore, 1D quantum mechanics with point interactions which belongs to a class of the quantum graph~\cite{Ohya:2015xya}. The systematic construction of such a Berry's connection is also discussed in \cite{Ohya:2020lzi}.
Since the general quantum graph has a wide parameter space, we can expect that the structure of the Nahm construction is also hidden and other BPS monopoles appear as the Berry's connection on the boundary condition.
These issues will be reported in future works.

\section*{Acknowledgements}

The authors thank H. Sonoda and S. Ohya for useful discussions.
This work is supported by Japan Society for the Promotion of Science (JSPS) KAKENHI Grant Number JP18K03649 (MS) and JP21J10331 (TI).

\small
\bibliographystyle{utphys}
\bibliography{ref}

\end{document}